\documentclass[usenatbib]{mn2e}
\usepackage{times}
\usepackage{epsfig}
\usepackage{colordvi}
\usepackage{graphicx}
\usepackage{amssymb,textcomp}

\newcommand{\apj}{ApJ}
\newcommand{\apjl}{ApJ}

\newcommand{\aap}{A\&A}

\newcommand{\mnras}{MNRAS}
\newcommand{\nat}{Nature}

\newcommand{\ssr}{Space Sci. Rev.}        
   
\newcommand{\msun}{\mbox{M}_{\sun}}    

\newcommand{\rmd}{{\rm d}}

\title[Reprocessing model for QPOs in  black holes]{Reprocessing model for the optical quasi-periodic oscillations
in black hole binaries}
 \author[Alexandra Veledina and Juri Poutanen]{Alexandra Veledina$^{1,2}$
 and Juri Poutanen$^{2}$\thanks{E-mail: juri.poutanen@utu.fi}   
  \\
$^1$Astronomy Division, Department of Physics, PO Box 3000, FI-90014 University of Oulu, Finland \\
$^2$Tuorla Observatory, Department of Physics and Astronomy, University of Turku, V\"ais\"al\"antie 20, FI-21500 Piikki\"o, Finland\\
}

\begin{document}

\volume{448}
\pagerange{939--945}
\pubyear{2015}

\date{Accepted 2014 December 22. Received 2014 November 26; in original form 2014 September 23}

\maketitle
\label{firstpage} 

\begin{abstract}
\noindent
A number of black hole X-ray transients show quasi-periodic oscillations (QPOs) in the optical (ultraviolet) and X-ray
bands at the same frequency, which challenge models for production of radiation at these wavelengths.
We propose a model where the optical radiation is modulated by the oscillating X-ray flux 
resulting in varying irradiation of the outer parts of the accretion disc. 
The proposed QPO mechanism inevitably takes place in the systems with sufficiently small 
ratio of the outer disc radius to the QPO period. 
We show that, unlike in the case of the aperiodic variability, it is not possible to obtain the optical
QPO profiles from those observed in the X-rays through the transfer function, 
because of different X-ray signals seen by the disc and by the observer.
We demonstrate that with the increasing QPO frequency,  occurring at the rising phase of the X-ray outburst, the rms 
should be constant for sufficiently low frequencies, then to increase reaching the peak and finally to drop substantially 
when the QPO period becomes comparable to the light-crossing time to the outer disc. 
We predict that the QPO rms in this model should  increase towards shorter wavelengths 
and this fact can be used to distinguish it from other QPO mechanisms.  
\end{abstract}



\section{Introduction}

The significant influence of the X-ray source on the structure and spectral properties 
of the standard accretion disc was noticed already in the pioneering study by \citet{SS73}
of the optically thick geometrically thin $\alpha$-discs. 
The temperature distribution in such irradiated discs around black holes (BHs) was calculated 
in a number of works \citep*[see e.g.][]{Cun76,FKR02}.
Though the precise dependence of the effective temperature on radius depends on the model,
the general expectation is that X-ray irradiation dominates over internal viscous dissipation 
in the outer part of the accretion disc, which consequently dominates the emission 
in the optical/infrared (OIR) wavelengths.

The prediction was then confirmed by the OIR luminosity and the specific shape of the OIR spectra in a 
number of BH binaries (\citealt{vPMC94}; \citealt*{GDP09};  \citealt{ZCC11};  \citealt*{CCZ12};  \citealt*{PVR14}).
The signature of reprocessed X-ray emission are also seen in optical/X-ray cross-correlation 
function \citep[CCF,][]{Hynes98,HBM09}, displaying a single, a few second broad peak at positive optical lags, 
as well as in the transfer function determined directly from the light curves \citep{OBrHH02}.
Additional support comes from the detection of optical flares following X-ray bursts 
\citep[e.g.][]{Grindlay78,McClintock79,Hynes06}. 
On the other hand, sometimes the complex shape of the CCFs displaying the so-called precognition dip 
cannot be explained solely by the irradiation of the disc \citep{Motch83,Kanbach01,DGS08,HBM09,GDD10}, 
an additional contribution of another source, likely of synchrotron origin, is required \citep*{MMF04,VPV11,PV14}.

\begin{figure*}
\epsfig{file=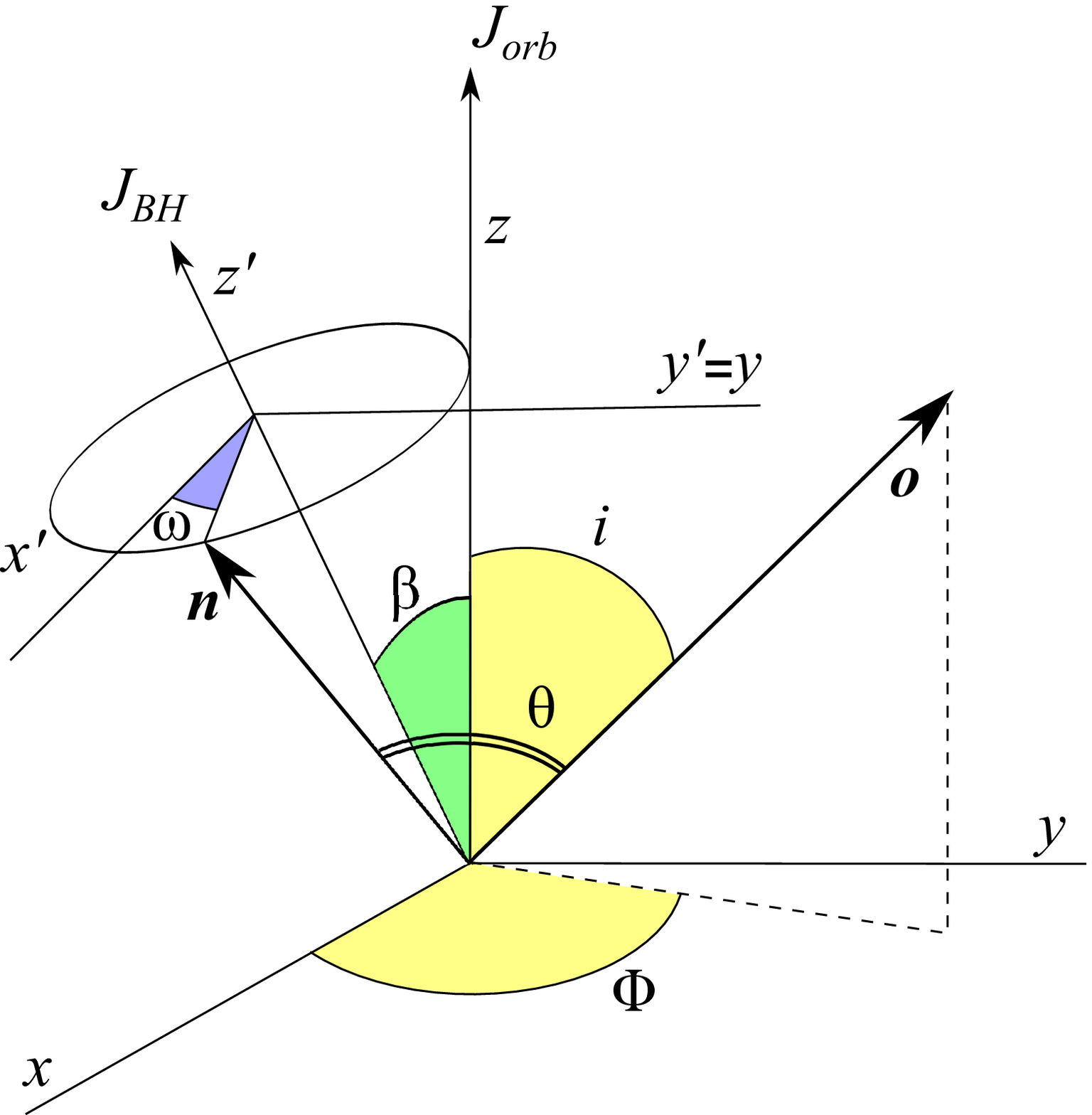, width=6cm} \hspace{0.5cm}
\epsfig{file=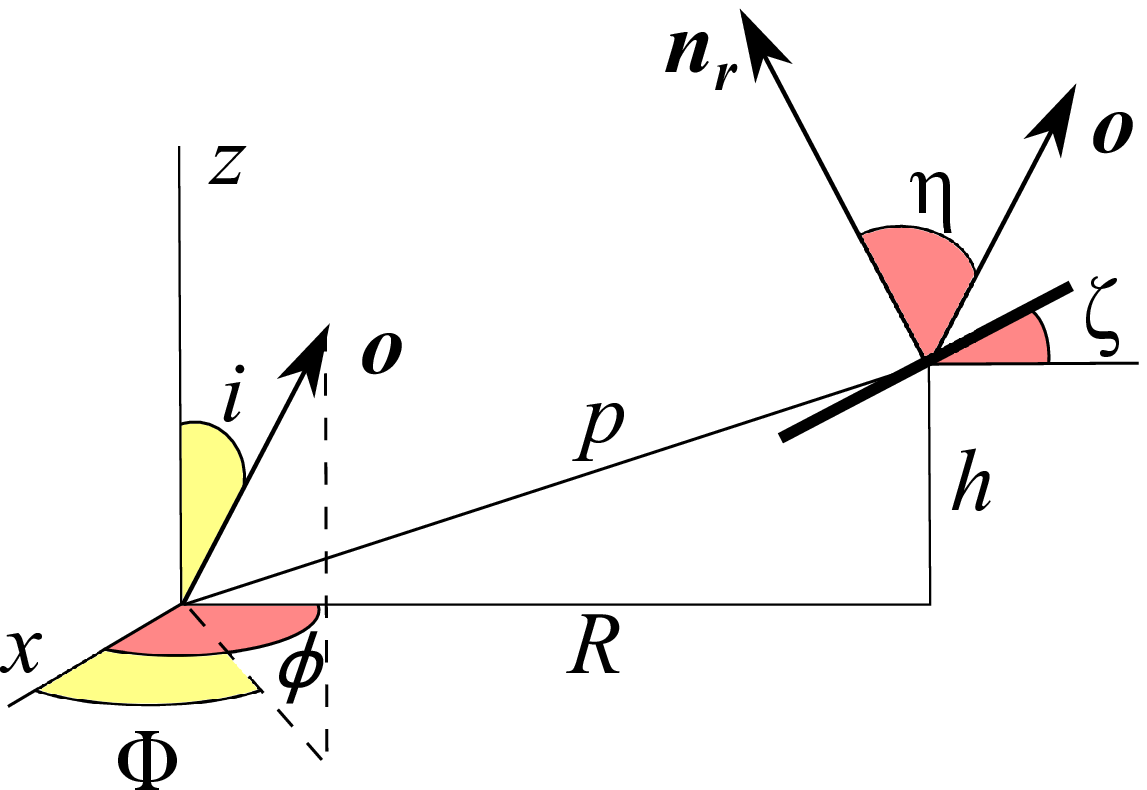, width=6cm}
\caption{
Schematic representation of the hot flow producing X-ray and optical QPOs.
Left: coordinate systems connected with the orbital plane $xyz$ and with the BH spin $x'y'z'$ are shown.
The plane $xy$ coincides with the orbital plane and $y$ is parallel to $y'$.
Axis $z'$ is aligned with the BH spin, which is inclined by the angle $\beta$ to the orbital spin.
The position of the observer $\bmath{o}$ is described by the azimuthal angle $\Phi$ and binary inclination $i$.
The position of the normal to the hot flow $\bmath{n}$ is characterized by the precession
angle/phase $\omega$.
It makes an angle $\theta$ with the direction to the observer, which depends on $\omega$.
Right: schematic representation of the reprocessing geometry. 
The surface element is at radius $r$ and height $h$ from the central X-ray source, its normal makes angle 
$\zeta$ with the orbital spin and angle $\eta$ with the line of sight. 
}\label{fig:geometry_flow} 
\end{figure*}

Simultaneous presence of at least two optical components is also required by the shape of the 
power spectral density (or, equivalently, the auto-correlation function): while the optical photons 
are thought to originate from a somewhat larger region compared to the X-rays, the power spectrum 
does not have the characteristic suppression of the high-frequency noise, instead, an 
additional Lorentzian peaking at higher frequencies (with respect to the X-rays) is required 
\citep{Kanbach01,GDD10}.
That Lorentzian might be a manifestation of the interplay of different optical components, 
one related to the synchrotron radiation from the hybrid hot flow and another to the irradiated disc 
\citep{VPV11}.

Apart from the broad-band noise, the quasi-periodic oscillations (QPOs) were detected in the optical 
\citep{Motch83,DGS09,GDD10} and UV \citep{HHC03} wavelengths at frequencies 0.05--0.13~Hz.
The X-ray power-density spectra demonstrate similar features known as the type-C low-frequency QPOs 
\citep[see][for classification]{CBS05}. 
The two months of observations revealed that the optical, UV and the X-ray QPOs in BH binary  
XTE~J1118+480 share a common (within uncertainties) characteristic frequency whilst evolving 
with time \citep{HHC03}.
This suggests that the QPOs in these three energy bands are parts of the same process.
One or several components responsible for the optical emission -- the hot accretion flow \citep*{VPV13}, 
the jet \citep[e.g.][]{HRP06} and the irradiated disc \citep{GDP09} -- are all natural candidates for the QPO origin.
Because the principal oscillation mechanisms and the geometrical properties are substantially 
different, these three models should be distinguishable by the observational characteristics.
The hot flow QPO model was considered in \citet*{VPI13}.
In this work, we develop a quantitative model for the low-frequency QPOs arising from reprocessing.

One of the most promising mechanisms to produce X-ray QPOs  is based on the misalignment of the accretion flow 
and the BH spin. The QPOs there arise due to the Lense--Thirring (solid-body) precession of the whole inner hot flow 
(\citealt{FB07};  \citealt*{IDF09,ID11}). 
Recalling that the cold accretion discs are likely to be flared \citep[e.g.][]{SS73,FKR02}, any temporal variations 
of the X-ray flux should be reflected in the reprocessed radiation.
In this paper we introduce a model for the optical QPOs produced by reprocessing of the X-ray modulated flux.  
We present detailed calculations of the QPO profiles and amplitudes. 
Even without any calculations we can predict that the optical QPOs have 
to be visible if the light-crossing time of the accretion disc is shorter than the QPO period. 
If this condition breaks down, QPOs still might be visible at  short wavelengths because 
emission there is produced closer to the irradiating source. 
We discuss then how the optical QPOs arising from the reprocessing can be distinguished from those produced by 
other QPO mechanisms (i.e., produced by the hot accretion flow or by the jet).

\section{Geometry and formalism}

\subsection{X-ray QPOs}
\label{sect:xray_qpos}

We consider a simple scenario of a flat precessing disc radiating in the X-ray band. 
We associate this geometry with the hot accretion flow around a Kerr BH, 
which undergoes solid-body Lense--Thirring precession because of the misalignment of the BH and orbital spins \citep{FB07,IDF09,ID11}. 
The general geometry is shown in Fig.~\ref{fig:geometry_flow} (left).
We define two coordinate systems: $xyz$ associated with the binary orbital plane 
and $x'y'z'$ tied to the BH spin with the  $z$-axes being aligned with the orbital spin and the BH spin, respectively.
The angle between them is denoted by $\beta$. We further choose axis $y=y'$.
The precession occurs around the BH spin axis with the precession angle $\omega$ 
measured from $x'$-axis, so that the instantaneous hot flow normal 
$\hat{\bmath{n}}$\footnote{We use a hat to denote unit vectors throughout.}
is aligned with the orbital spin axis at $\omega=\upi$ and has a maximal misalignment of 
$2\beta$ when $\omega=0$.
The observer position is described by
$\hat{\bmath{o}}=(\sin i \cos\Phi, \sin i \sin\Phi, \cos i)$ in $xyz$ coordinates, where
$i$ is the binary inclination and $\Phi$ is the azimuth of the observer measured from the $x$-axis.
The hot flow normal is given by $\hat{\bmath{n}}=(\sin\beta \cos\omega , \sin\beta \sin\omega, \cos\beta)$ 
in $x'y'z'$ coordinates.
It translates to 
\begin{equation}\label{eq:disc_normal}
\hat{\bmath{n}} = \left(\sin\beta \cos\beta(1+\cos\omega),  \sin\beta \sin\omega, 1-\sin^2\beta (1+\cos\omega)\right) 
\end{equation}
in $xyz$ coordinates.

The X-ray QPOs arise from the different orientation of the hot flow relative to the observer, 
described by the scalar product $\hat{\bmath{o}} \cdot \hat{\bmath{n}} =\cos\theta$.
It is easy to show that
\begin{eqnarray}\label{eq:cos_theta_obs}
   \cos\theta &=& \sin\beta \cos\beta \sin i  \cos\Phi\  (1+\cos\omega)   \\ 
           &+& \sin\beta \sin\omega \sin i  \sin\Phi + \cos  i\ [1-(1+\cos\omega)\sin^2\beta]. \nonumber
\end{eqnarray}
The flux observed from one ring of the flat disc with radius $r$, thickness $\d r$ and a surface normal which
makes an angle $\theta$ to the line of sight far from the BH can be expressed as
\begin{equation}\label{eq:flux_azimuth_ave2}
 \rmd F_E (r, \theta) = \frac{r\ \rmd r}{D^2}  q_E(r) f_E(r,\theta).
 \end{equation}
Here, $D$ is the distance to the observer, $q_E(r)$ is the surface flux per energy interval at a given radius and the factor
$f_E(r,\theta)$ accounts for the angular dependence of the observed flux and is calculated from the specific intensity 
emerging from a surface element.
In general, the specific intensity depends on the zenith angle. 
X-ray emission from accreting BHs is produced by Comptonization in an optically translucent flow with Thomson optical depth $\tau\sim1$.
In this case, the local angular dependence of the radiation intensity  in the frame comoving with the considered element 
can approximately be described by \citep{PG03}
\begin{equation}\label{eq:compton_pattern}
 I_{E}(\mu) \propto 1 + b \mu ,
\end{equation}
where  $\mu$ is the cosine of the zenith angle and the parameter $b\approx -0.7$ (for exact solutions see \citealt{ST85,VP04}).
We assume that the emergent spectrum from all surface elements is a power law with  photon index $\Gamma=1.7$.
To calculate the observed flux, we take into account gravitational redshift, Doppler shift, time dilation and light bending 
in the Schwarzschild metric\footnote{This gives very similar results to using the more appropriate Kerr metric for the $r>3R_{\rm S}$ 
case we consider here.} following techniques presented in \citet{PG03} and \citet{PB06}.
In the absence of the standard energy dissipation profile relevant to the hot flow in general relativity, we take the standard 
profile of a thin disc \citep{SS73}
\begin{equation}\label{eq:SS73_profile}
q_E(r) \propto \sqrt{1-\frac{3R_{\rm S}}{r}}  r ^{-3},
\end{equation}
which is suitable for illustration. Here $R_{\rm S}=2GM/c^2$ is the Schwarzschild radius of the BH. 
We define the weighted angular emissivity function as 
 \begin{equation}\label{eq:wei_ftheta}
\overline{f}_E(\theta) = \frac{\int  f_E(r,\theta)\ q_E(r)\ r\ \rmd r } {\int q_E(r)\ r\ \rmd r }, 
\end{equation}
where the integration is performed over the hot flow surface with the radius 
varying  between 3 and 100$R_{\rm S}$.
The observed flux is 
\begin{equation}\label{eq:flux_obs}
F_E = \frac{4\upi \int q_E(r)\ r\ \rmd r }{4\upi\ D^2} \   \overline{f}_E(\theta) .
 \end{equation}
The angular dependence of the X-ray flux is shown in fig.~3 of \citet{VPI13}, 
where  further details on the precessing hot flow model can be found.

\subsection{Reprocessing model}
\label{sect:models_setup}
 
In the reprocessing model, the emission is directly related to the central source X-ray flux 
that shines upon the cold disc. 
Here we follow formalism developed by \citet{Pou02} for X-ray reflection from a flared disc. 
We assume that the reprocessing into optical wavelengths occurs in a ring extending from radius $R_{\min}$ 
to $R_{\max}$,  large compared to the extent of the X-ray source which therefore can be
considered as an anisotropic point source.
Depending on the wavelength of interest, $R_{\max}$ can be equal to or smaller than the disc outer radius $R_{\rm disc}$.
We assume the power-law dependence of height on radius $h= H(r/R_{\rm disc})^\rho$.
The parameters describing the disc shape are the ratio $H/R_{\rm disc}$ and 
the power-law index $\rho$.  
The ring surface makes an angle $\zeta$ with the orbital plane that is a function of $r$: 
\begin{equation}
 \tan \zeta = \rho \frac{h}{r}= \rho \frac{H}{R_{\rm disc}} \left( \frac{r}{R_{\rm disc}} \right)^{\rho-1}
\end{equation}
(see Fig.~\ref{fig:geometry_flow}, right). 
The standard accretion disc has  $\rho=9/8$ \citep{SS73}, while the irradiated disc 
may have $\rho=9/7$ \citep{FKR02}, thus we only consider cases with $\rho>1$. 
In the frame related to the orbital plane,  the observer's coordinates are 
\begin{equation}
\hat{\bmath{o}}=(\sin i \cos\Phi, \sin i \sin\Phi, \cos i), 
\end{equation}
the radius-vector $\bmath{p}$ pointing towards a surface element has coordinates
\begin{equation}\label{eq:p_vector}
 \bmath{p} = (r\cos\phi, r \sin\phi, h), 
\end{equation}
and the normal to the element is 
\begin{equation}
 \hat{\bmath{n}}_{\rm r}=(-\sin\zeta \cos\phi, -\sin\zeta \sin\phi, \cos\zeta) . 
\end{equation}

The optical light curve  from such a ring can generally be written as
\begin{eqnarray} \label{eq:repro_lum}
 L_{\rm repr} (t) \propto \! \!\!   \int \limits_{-\infty}^{t} \! \!\!  \rmd t'  \! \! 
\! \!\!  \int \limits_{R_{\min}}^{R_{\max}} \! \!  \frac{ r\,\rmd r \cos \xi }{4\upi p^2 \cos\zeta}  
  \int \limits_{0}^{2\upi} \! \! \rmd\phi  L^{\alpha}_{\rm X} (t',  \bmath{p}) \delta(t\! -\! t'\! -\! \Delta t(\phi)) {\cos \eta}. 
\end{eqnarray}
Here  $L_{\rm X} (t',  \bmath{p})$ is the X-ray luminosity in the direction 
of the disc element, i.e. at angle  $\theta_{\rm e} = \arccos(\hat{\bmath{n}}\cdot \hat{\bmath{p}})$ relative to the hot flow normal,
emitted from the central source at time $t'$. 
The index $\alpha$ may depend on the wavelength where reprocessing signal is measured. 
For example, assuming that disc radiation is a blackbody of some temperature $T$
and remembering that $T \propto L_{\rm X}^{1/4}$, it is obvious that 
in the Rayleigh--Jeans part of the spectrum $\alpha\approx 1/4$, 
while closer to the peak of the reprocessed emission, in the UV range, $\alpha$ is close to unity 
or may even exceed 1 in the Wien tail of the spectrum. 
In reality, of course, for a given wavelength $\alpha$ varies, because of the varying 
temperature of the irradiated disc. Here, we ignore this effect and concentrate on the 
geometrical factors which are dominant.

The $\delta$-function accounts for the geometrical time-delays: 
\begin{eqnarray}
 \Delta t(\phi)&=&\frac 1c (p - {\bmath p} \cdot {\hat{\bmath o}} ) \\
               &=&\frac 1c \left[\sqrt{r^2+h^2}-h\cos i -r \sin i\cos(\Phi-\phi) \right]. \nonumber
\end{eqnarray}
We neglect the reprocessing time, which is orders of magnitude smaller than 
any other time-scales considered here.
The area of the surface element is $r\ \rmd r\ \rmd\phi/\cos\zeta$ and 
the projection of this element on the line connecting it to the X-ray source is proportional to 
\begin{equation}
\cos\xi = - \hat{\bmath{n}}_{\rm r} \cdot \hat{\bmath{p}}= \frac{(\rho-1)h/r}{\sqrt{1+(h/r)^2}}\cos\zeta .
\end{equation}

Let us assume for simplicity that the angular distribution of reprocessed radiation follows Lambert law, implying 
that the luminosity of the disc element per unit solid angle depends linearly on 
its projection on the observer's sky, which is proportional to 
\begin{equation}
\cos\eta = \hat{\bmath{n}}_{\rm r} \cdot \hat{\bmath{o}}=\cos i \cos\zeta - \sin i \sin\zeta \cos(\Phi-\phi) .
\end{equation}
 
Using the $\delta$-function to take the integral over $t'$ and rewriting the light curve 
in terms of the QPO phases of the  signal corresponding to the arrival and emission times 
$\omega=2\upi \nu_{\rm QPO} t$ and $\omega'=2\upi \nu_{\rm QPO} t'$, we  get: 
\begin{equation} \label{eq:reprocessing}
L_{\rm repr} (\omega) \propto 
\int \limits_{R_{\min}}^{R_{\max}} 
                     \frac{\rmd r}{2\upi r} \frac{(\rho-1) h/r}{(1+(h/r)^2)^{3/2}}  
                     \int \limits_{0}^{2\upi}  \rmd\phi \ \cos\eta\ 
                    \overline{f}^{\alpha}_{\rm X}(\cos\theta_{\rm e})    ,
\end{equation}
where  the X-ray flux $\overline{f}_{\rm X}$ in the direction of the elements depends on the angle 
between the hot flow normal $\hat{\bmath{n}}$ and vector $\bmath{p}$
(see equations (\ref{eq:disc_normal}) and (\ref{eq:p_vector})): 
\begin{eqnarray}\label{eq:cos_theta}
\cos\theta_{\rm e}&=& \hat{\bmath{n}}\cdot \hat{\bmath{p}}=  
                      \frac{1}{\displaystyle \sqrt{1+(h/r)^2}} \left\{\left[1\! - \! (1+\cos\omega') \sin^2\beta\right] \frac{h}{r} \right. \nonumber \\
                  &+&\! \!\!  \! \!  \left.\cos\phi  \sin\beta \cos\beta (1+\cos\omega' )\!  +\!   \sin\phi \sin\beta \sin\omega' \right\}, 
\end{eqnarray}
where 
\begin{equation}\label{eq:ww'}
 \omega'= \omega - \frac{2\upi  \nu_{\rm QPO}}{c}\left[ \sqrt{r^2+h^2} - h \cos i - r\sin i \cos(\Phi-\phi)\right] .
\end{equation}
 
For the disc surface element at radius $r$ and azimuth $\phi$ to contribute to the observed flux, 
three conditions have to be satisfied:
\begin{enumerate} 
\item the irradiation condition (the X-ray source is seen from the ring surface element), i.e. $\cos\theta_{\rm e}>0$; 
\item the visibility condition (the observer sees the ring surface), i.e. $\cos\eta >0 $ or, equivalently, $\cos(\Phi-\phi)<\cot i \cot\zeta$;  
\item the reprocessed photons are not blocked by the disc on their way to the observer, i.e. 
\begin{equation} \label{eq:irr_cond3}
\tan i < \frac{\sqrt{R_{\rm disc}^2-r^2\sin^2(\Phi-\phi)}-r\cos(\Phi-\phi)}{H-h}. 
\end{equation}  
\end{enumerate}

The model parameters can be divided into two groups: the disc parameters $\rho$, $H/R_{\rm disc}$, $R_{\min}/R_{\max}$, 
$R_{\max}/R_{\rm disc}$, 
$\nu_{\rm QPO}R_{\max}/c$, and $\alpha$ and the orientation parameters $i$, $\Phi$ and $\beta$.
The parameters $\alpha$ and $\beta$  
only affect the signal rms, rather than the profile shape.
The first three parameters also make rather minor changes to the QPO signal as we discuss below. 
The fourth parameter is equal to unity in the OIR wavelengths, where the reprocessed radiation is dominated by the outer parts of the disc, 
and for the sake of simplicity in the following we take $R_{\max}=R_{\rm disc}$. 
Thus the model in reality has only three main parameters: $\nu_{\rm QPO}R_{\max}/c$, which controls the smearing of the 
X-ray QPO signal in the disc, and the orientation parameters $i$ and $\Phi$.

\begin{figure*}
\centerline{\epsfig{file=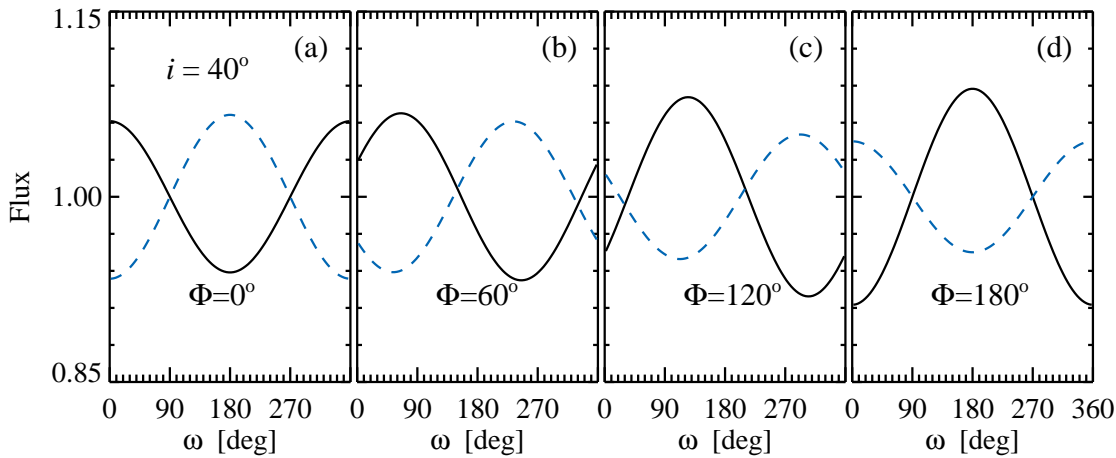, width=15cm}}
\caption{ 
The X-ray (black solid line) and computed optical (blue dashed line) QPO profiles for the case $\nu_{\rm QPO}R_{\max}/c=10^{-5}$, 
$i=40\degr$, $\beta=5\degr$, $H/R_{\max}=0.2$, $\rho=1.5$, $R_{\min}/R_{\max}=0.99$ and $\alpha=1$
and four different observer's azimuthal angles $\Phi$ are considered (panels a--d).
Here we assume the X-ray emissivity function $\overline{f}_{\rm X}\propto\cos\theta_{\rm e}$.
The profiles are in excellent agreement with the analytical solution (\ref{eq:lrepr_analyt}).
} \label{fig:analytic_profile}
\end{figure*}

\section{Results}

\subsection{QPO profiles} 

In contrast to the model with QPOs from the hot flow \citep{VPI13}, here optical profiles depend on the QPO frequency.
If the QPO period is much shorter than the corresponding light-travel time to the outer disc
($\nu_{\rm QPO}R_{\max}/c\gg1$), 
the variability amplitude of the optical QPO should go to zero.
For very long QPO periods, $\nu_{\rm QPO}R_{\max}/c\ll 1$, the delays due to light-travel time 
can be neglected and  the QPO profiles just reflect  variations of the illumination of the disc by the X-ray source and 
of the viewing angle of the disc surface.
For the parameters $R_{\rm disc} \sim 10^{10} - 10^{12}$~cm (typical for low-mass X-ray binaries) 
and the QPO range between $10^{-3}\lesssim\nu_{\rm QPO}\lesssim10$~Hz,
the possible range of parameter $\nu_{\rm QPO}R_{\max}/c$ is between $\sim10^{-4}$ and $300$.
It is therefore clear that the limiting cases discussed above can, in principle, be realized.

The integrals in equation~(\ref{eq:reprocessing}) should generally be computed numerically; however, it is interesting 
to consider a case when they can be taken analytically.  
This can also be used as a benchmark for the precise numerical calculations.
Let us first consider the situation where the reprocessing occurs in a thin ring ($R_{\min}\cong R_{\max}$), which is located 
sufficiently close to the X-ray source ($\nu_{\rm QPO}R_{\max}/c\ll1$), i.e. the response is immediate and we can put 
$\omega=\omega'$.
In addition, the ring is located high enough above the orbital plane and the angle $\beta$ is small enough so that the 
X-ray source is always seen from the entire ring (this translates to the condition $\cos\theta_{\rm e}>0$ at any phase).
We assume that the emission pattern of the central source follows the Lambert law, 
which leads to the emissivity function $\overline{f}_{\rm X}\propto\cos\theta_{\rm e}$ (we also put $\alpha=1$).
Under these conditions, the integral in equation~(\ref{eq:reprocessing}) is reduced to the simple 
expression
\begin{equation} \label{eq:lrepr_analyt}
  L_{\rm repr} (\omega) \propto 1-\sin^2\!\beta\, (1 + \cos\omega) - \frac{\rho}{2}\tan i \sin\beta \left[ \cos\Phi + \cos(\Phi-\omega)\right].
\end{equation}
As an illustration, we plot the resulting profiles for the aforementioned conditions in Fig.~\ref{fig:analytic_profile}.
Interestingly, the reprocessed emission is coming (almost) in anti-phase with the X-rays.
For $\beta\ll1$ it is easy to see from equation~(\ref{eq:lrepr_analyt}) that the minima and maxima are achieved at 
$\omega=\Phi$ and $\Phi+\upi$, respectively, and vice versa for the X-rays (see equation~(\ref{eq:cos_theta_obs})).
Physically, the maximum in reprocessing is achieved when the X-rays are illuminating the opposite from the observer 
side of the disc which has the largest projected area.
This analytical solution, though, is unlikely to be realized in a realistic situation,
as the X-ray emissivity function deviates from the simple cosine dependence.

\begin{figure*}
\centerline{\epsfig{file=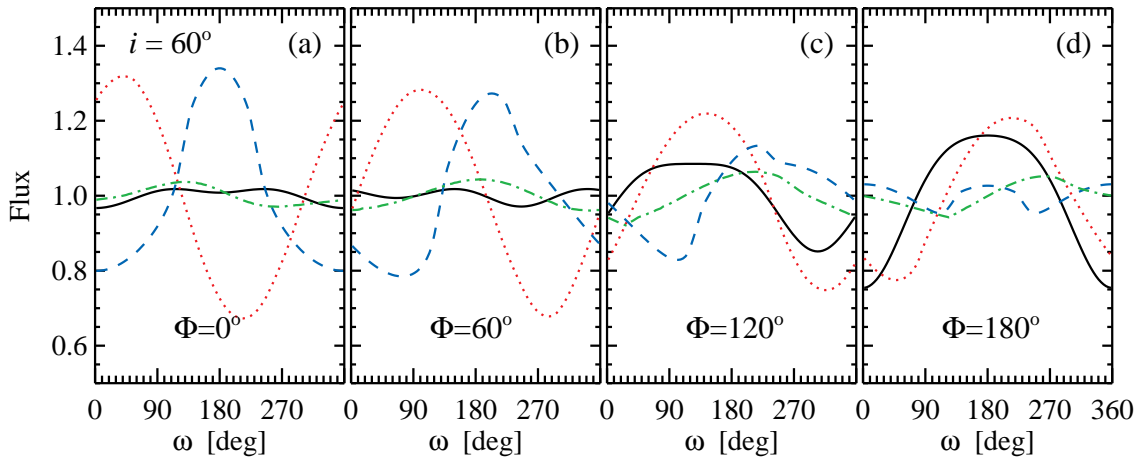, width=15cm}}
\caption{ 
Possible QPO profiles for the three cases of the parameter $\nu_{\rm QPO}R_{\max}/c$: $10^{-2}$ (blue dashed), 0.5 (red dotted) and 
2 (green dot--dashed).
The system inclination is fixed $i=60\degr$ and four different observer's azimuthal angles $\Phi$ are considered (panels a--d).
X-ray profiles are calculated for the full hot flow extending from $3$ to $100R_{\rm S}$ (black solid  lines) assuming BH spin inclination of $\beta=10\degr$. 
Other parameters of the reprocessing model are $R_{\min}/R_{\max}=0.5$,  $H/R_{\max}=0.2$, $\rho=1.5$ and $\alpha=1$.
} \label{fig:reproc_profile_routnuqpo}
\end{figure*}

We investigate the effect of changing parameters on the QPO profiles by the direct calculations of reprocessing light curves
for a more realistic X-ray emissivity function (see Section~\ref{sect:xray_qpos}).
Optical and X-ray QPO profiles are calculated by assuming a radial emissivity $q_E(r)$ for each band. 
We fix $\beta=10\degr$ and an observer inclination $i=60\degr$ and consider changing $\rho$, $H/R_{\max}$, $R_{\min}/R_{\max}$
and $\nu_{\rm QPO}R_{\max}/c$.
We find that the parameter $\rho$ does not alter the profile shapes and only affects the amplitudes of oscillations 
(with higher rms obtained for the cases when the observer sees relatively larger disc area), while the decrease of 
$R_{\min}/R_{\max}$ results in somewhat faster response and more smearing of the signal, i.e. lower rms.
The increase of parameter $H/R_{\max}$ generally introduces a larger phase shift of the optical light-curve, in addition, 
the larger is this parameter, the lower is the optical variability rms.
This results from the decrease of the X-ray rms towards lower inclinations (or, essentially, towards lower angles $\theta_{\rm e}$).
The major effect on the QPO profiles comes from changing parameter $\nu_{\rm QPO}R_{\max}/c$.
We consider three cases: $\nu_{\rm QPO}R_{\max}/c=10^{-2}$, 0.5 and 2.
The resulting profiles are shown in Fig.~\ref{fig:reproc_profile_routnuqpo}.

It is evident that the optical QPOs arising from reprocessing have larger amplitudes than those seen 
in the X-rays by the distant observer (with their ratio reaching a factor of 10 at low azimuth $\Phi<60\degr$). 
We have studied the QPO profiles at the $i$--$\Phi$ plane and found that this is true for majority of the simulations (except for the area
with $i\gtrsim70\degr$ and $\Phi\gtrsim120\degr$, which, however, cannot be described properly with the presented 
model, because the effects of non-zero thickness of the hot flow become important).
The main reason for that is the strong increase of the rms in X-rays with increasing viewing angle 
(i.e., inclination in the case of observer and the corresponding angle $\upi/2-\xi$ for the disc). 
Small opening angles (realistic values are between 6\degr\ and 22\degr, see \citealt{deJong96} and references therein) thus result in 
substantially higher variation amplitudes of the X-rays seen by the disc relative to those seen by the observer.
In addition, the X-ray variability seen by the disc is different from that seen by the observer for the same reason.
This means that it is not feasible to find any kind of QPO response function, which could give the optical QPO profiles 
from the observed X-ray QPOs.

Another interesting observation is that at  $\nu_{\rm QPO}\ll  c/R_{\max}$ the 
optical QPO profile strongly depends on $\Phi$, being nearly sinusoidal at small $\Phi$, 
with the growing  harmonic content at larger $\Phi$, where the amplitude of the fundamental drops significantly. 
The reasons for that being the violation of the irradiation condition (\ref{eq:irr_cond3}) and the complex shape of X-ray emission pattern.
At higher $\nu_{\rm QPO}\sim c/R_{\max}$, the profiles are more sinusoidal with a low  harmonic content. 
 
We finally note that the overall optical rms can be reduced by the presence of other, non-oscillating or oscillating out-of-phase, components.
For instance, additional constant flux may arise because of the viscous heating in the disc.
Also, the X-ray source likely has a non-zero thickness, thus the entire disc has an additional constant irradiated component.
And finally, reflection of the X-rays from the optically thin corona or wind would produce a nearly constant illumination 
over the disc surface leading to the reduction of the QPO amplitude.

 \begin{figure*}
\centerline{\epsfig{file=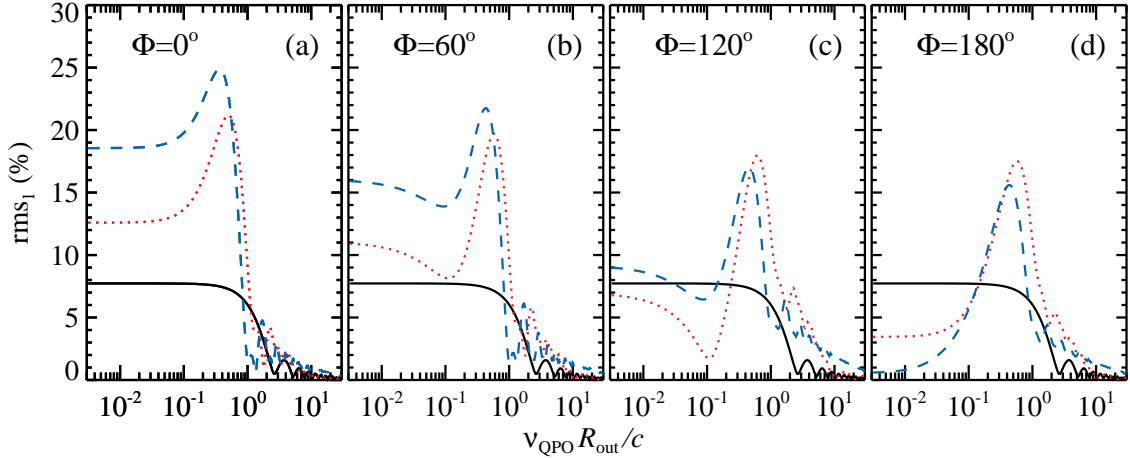, width=15cm}}
\caption{ 
The dependence of the rms at the fundamental frequency on $\nu_{\rm QPO}R_{\max}/c$ 
for different observer's azimuth $\Phi=0, 60, 120$ and 180\degr (panels a-d, respectively) 
and three different inclinations, $i=0, 40\degr$ and 60\degr shown by solid black, dotted red and dashed blue curves. 
} \label{fig:rmsnuqpo}
\end{figure*}
x

\subsection{QPO amplitude dependence on Fourier frequency} 
 
Often the optical QPO profile cannot be measured and the only 
available information is its rms.  
The ratio of the light-crossing time to the outer disc to the QPO period,  $\nu_{\rm QPO}R_{\max}/c$, is 
the major factor affecting the rms, see Fig.~\ref{fig:rmsnuqpo}. 
The constant level at low frequencies reflects the fact that the light-travel time to the outer disc is much 
below the QPO period, so that the response can be considered as immediate. The QPO profile is 
then completely determined by the varying illumination of the outer disc due to precession of the 
anisotropic and non-axisymmetric X-ray source.
We see that the constant level is very different for different  observer's azimuth $\Phi$, 
being close to 20 per cent at $\Phi\sim0\degr$ and dropping to nearly zero at $\Phi\sim 180\degr$ 
(we note that the harmonic here is stronger). 

At high $\nu_{\rm QPO}$, the rms drops dramatically due to smearing. 
Rather unexpectedly we find that the rms has  a strong peak at $\nu_{\rm QPO}R_{\max}/c\sim1/2$.  
The peak does not appear only for inclination $i=0$. 
For $\Phi\lesssim 60\degr$, the rms reaches 17--25 per cent, depending on the inclination. 
At larger $\Phi \gtrsim 120\degr$, the rms is about 15 per cent which is substantially larger than that at low $\nu_{\rm QPO}$. 
This resonance-like feature results from the fact that the reprocessing signals 
from both the closest and the furthest  (to the observer) parts of the disc are coming in phase.

We conclude here that the optical QPOs arising from  reprocessing of the X-ray radiation from an 
anisotropic precessing source can be identified in the data using its frequency dependence.
The rms is predicted to be constant at low frequencies, to have a peak at $\nu_{\rm QPO}R_{\max}/c\sim1/2$ 
and disappear at $\nu_{\rm QPO,\max}R_{\max}/c \sim 1$.  
The critical QPO frequency  depends on the disc size, which is itself a function of binary separation, 
$R_{\rm disc}\approx 0.6 a$ (for small mass ratios; see \citealt{Warner95}). Using the third Kepler law 
(and taking $R_{\rm disc}=R_{\max}$), we can relate 
it to the orbital period (in hours): 
\begin{equation}\label{eq:numax}
\nu_{\rm QPO,\max}\approx \frac{2}{3} P_{\rm h}^{-2/3} (M/10\msun)^{-1/3}\ \mbox{Hz}.
\end{equation}

\subsection{QPO amplitude  dependence on the wavelength} 

An alternative way to identify the reprocessing QPO in the data is from its wavelength dependence. 
Different wavelengths correspond to different parts of the irradiated disc (blackbody) spectrum, resulting in 
variation of parameter $\alpha$ (introduced in equation~\ref{eq:repro_lum}). 
Combining relations $T\propto L_{\rm X}^{1/4}$ and $B_{\nu}(T)\propto L_{\rm X}^{\alpha(\nu)}$,
we get  
\begin{equation}
 \alpha(\nu) = \frac 14 \frac{\upartial \ln B_{\nu}(T)}{\upartial \ln T} 
             = \frac{h\nu}{4kT} \left( 1 - {\rm e}^{-h\nu/kT} \right)^{-1}.
\end{equation} 
In this paper, we considered cases with the linear response, $\alpha=1$, 
which takes place close to the wavelengths of the blackbody peak.
For the irradiated disc temperature of about 10\,000~K, this corresponds to the $U$-filter.
The rms is higher at shorter wavelengths, where $\alpha$ exceeds 1 and it is smaller at longer wavelengths, 
where it tends to the Rayleigh--Jeans value $\alpha=1/4$. 
Thus at long wavelengths the optical (or rather IR) QPO should have four times smaller rms than 
those computed here. On the other hand, in the Wien part of the spectrum $\alpha$ can exceed unity and 
the QPO amplitude grows. 

In addition to the wavelength dependence of the reprocessing QPO itself, 
it is likely that additional spectral components influence that. 
The irradiated disc spectrum may be  contaminated by the emission from the hot flow, jet, or circumstellar dust,  
which likely reduce the overall rms, with a somewhat larger suppression at longer wavelengths because 
their spectra are significantly redder than the spectrum of the disc. 
This fact further strengthens the dependence of rms on the wavelength, which should grow 
towards shorter wavelengths.

\subsection{Comparison with the data}

The low-frequency QPO range observed in the X-rays is $\sim10^{-2}-10$~Hz \citep{BS14}. 
For relatively small discs present in systems with periods of a few hours, the temperature at
the outer edge of the disc is substantially higher, roughly
$T_{\rm out} \propto F_{\rm X}^{1/4} \propto R_{\rm disc}^{-1/2} \propto P_{\rm orb}^{-1/3}$, where 
$P_{\rm orb}$ is the orbital period (we assumed that the X-ray luminosity is the same).
Thus, irradiation gives significant contribution to the UV flux, but not to the IR, where the disc is 
likely very dim compared to other components.
The discs in the long-period systems are large enough and cool enough to give significant contribution to
the IR, but the QPOs in such systems are expected to be seen only at low frequencies.
At the same time, the UV reprocessed emission in these systems is produced closer to the BH and hence the QPOs 
may be observed at relatively higher frequencies.
To calculate the profiles in this case, we would need to take into account 
the fact that $R_{\max}$ and the outer disc radius $R_{\rm disc}$ are not equal.

In GX~339--4, the optical QPO frequency was seen to vary between 0.05 and 0.13~Hz \citep{Motch83,Imam90,SteiCam97,GDD10}.
The QPO features were observed in three filters simultaneously \citep{GDD10}. 
They have systematically larger rms in the redder filters for all three nights of observations 
(see their table~4), though only the difference in rms observed during the third night are statistically significant.
The system period is rather long, about 42~h \citep{CJ14}, so that 
the maximum reprocessing QPO (see equation~(\ref{eq:numax})) is expected to be at $\nu_{\rm QPO,\max}\approx 0.055$~Hz, 
where we assumed $M=10\msun$. 
Thus the observed QPOs are likely produced either by the hot flow or in the jet. 

SWIFT~J1753.5--0127 is one of the shortest-period systems with $P_{\rm orb}\approx 3$~h \citep{ZDT08,Neustroev14}.
The optical QPOs in this object were detected during the simultaneous X-ray exposure 
at frequencies $\sim$0.08~Hz \citep{DGS09}.
The computed optical/X-ray CCFs do not resemble those of the irradiated disc, on the 
contrary, a prominent optical precognition dip is observed.
The dip may be a signature of the hot flow synchrotron emission \citep{VPV11}, and its amplitude relative to 
the positive peak (attributed to irradiation) implies that the hot flow is a dominant source of optical photons, 
and hence the QPOs. 

Both optical and UV QPOs were detected in XTE~J1118+480 at frequencies $\sim$0.07--0.13~Hz \citep{HHC03}.
The UV rms seems to be increasing with QPO frequency (see table~4 of \citealt{HHC03}), but because the errors on the
deduced rms are not listed, it is not clear whether this effect is real.
The system is otherwise a good candidate to expect the QPOs from irradiation, as it has a rather short period of 
about 4~h \citep{CJ14}.
Though the system demonstrates the precognition dip structure in the CCFs \citep{Kanbach01}, the dip amplitude is 
significantly smaller than that of the peak in the optical/X-ray CCF and the dip in the UV/X-ray CCF is even less 
prominent \citep{HHC03}.

\section{Summary}
\label{sect:summar}

We developed a model for the optical QPOs arising from the irradiation of the accretion disc by varying X-ray flux. 
Assuming that the inner part of the accretion disc is occupied by a precessing hot accretion flow (radiating in the X-rays) with the 
prescribed emission pattern, we calculate the optical profiles.
The oscillations at the precession period appear due to the changing of the illumination conditions.
We note that the distant observer sees the X-ray source at an angle different from that seen from the outer disc.
This means that simple use of the disc transfer function relating the observed X-ray and optical light curves 
is insufficient to reproduce optical profiles -- the additional 
knowledge of the X-ray emission pattern is required.

We find that the QPO rms is nearly constant for sufficiently low frequencies, $\nu_{\rm QPO}R_{\max}/c\lesssim0.1$, 
it then increases to achieve the maximum at $\nu_{\rm QPO}R_{\max}/c\sim1/2$ 
and drops dramatically at higher frequencies.
This is an identifying feature of the proposed mechanism, by which it can be distinguished from the optical QPOs 
arising from the hot flow.
Another way to recognize it is the dependence of the rms on the wavelength, where the increase of the variability 
amplitude towards shorter wavelengths is expected.

The described QPO mechanism inevitably should play a role in all systems with the sufficiently small 
ratio of the outer disc size to the QPO period. 
However, the observed oscillation amplitudes are expected to be lower than predicted here because 
of the substantial non-variable background. 
As other components, which do potentially  contribute to the optical emission (e.g. the hot flow and the jet), 
can also generate QPOs, finding the signatures of reprocessed emission in the oscillating signal becomes a challenging task.
The most promising low-mass X-ray binary systems, which are expected to have optical QPO 
due to reprocessing, are those showing a prominent positive peak in the optical/X-ray CCF,
corresponding to the irradiation delays of the aperiodic X-ray variability.
It is also preferable to search for these types of QPOs in the UV, where the disc radiation dominates and 
the QPO rms reaches the maximum, while other components are relatively dim.

\section*{Acknowledgements}

We thank Andrzej Zdziarski and Chris Done  for  useful suggestions.
The work was supported by the Finnish Doctoral Programme in Astronomy and Space Physics (AV) 
and  by the Academy of Finland grant 268740 (JP).


\label{lastpage}

\end{document}